\documentclass[showpacs, pra,onecolumn,preprintnumbers ,amsmath, amssymb, superscriptaddress, aps]{revtex4-2}
\usepackage{color}
\usepackage{amsmath,amssymb}
\usepackage{pifont}
\usepackage{amssymb}  
\usepackage{bbold}
\usepackage{float}
\usepackage{subfloat}

\usepackage[caption=false]{subfig}
\usepackage{tikz}
\usepackage{makecell}
\usepackage{subfig}
\usepackage{pifont}   
\usepackage{graphicx} 
\graphicspath{{Figures/}}
\usepackage{dcolumn}  
\usepackage{bm}       
\usepackage{multirow} 
\usepackage{placeins}
\usepackage[colorlinks]{hyperref}
\usepackage{mathtools}
\usepackage{appendix}

\def \be{\begin{align}}
	\def \ee{\end{align}}
\def \bea{\begin{eqnarray}}
	\def \eea{\end{eqnarray}}

\begin{document}

        \title{Magnetic field effect on tunneling through triple barrier in AB bilayer graphene}
        \date{\today}
        \author{Mouhamadou Hassane Saley }
        \email{hassmosy@gmail.com}
        \affiliation{Laboratory of Theoretical Physics, Faculty of Sciences, Choua\"ib Doukkali University, PO Box 20, 24000 El Jadida, Morocco}
        \author{Ahmed Jellal}
        \email{a.jellal@ucd.ac.ma}
        \affiliation{Laboratory of Theoretical Physics, Faculty of Sciences, Choua\"ib Doukkali University, PO Box 20, 24000 El Jadida, Morocco}
        \affiliation{Canadian Quantum Research Center, 204-3002 32 Ave Vernon,  BC V1T 2L7, Canada}

        
        \begin{abstract} 
We investigate electron tunneling in AB bilayer graphene through a triple electrostatic barrier {of heights $U_i (i=2,3,4)$ subjected to} a perpendicular magnetic field. By way of the transfer matrix method and using the continuity conditions at the different interfaces, the transmission probability is determined. Additional resonances appear for two-band tunneling at normal incidence, and their number is proportional to the value of $U_4$ {in the case of} $U_2<U_4$. However, when $U_2>U_4$, anti-Klein tunneling increases with $U_2$. The transmission probability exhibits an interesting oscillatory behavior when $U_3>U_2=U_4$ and $U_3 <U_2=U_4$. For fixed energy $E=0.39\gamma_1$, increasing barrier widths increases the number of oscillations and decreases Klein tunneling. The interlayer bias creates a gap for $U_2<U_3<U_4$ and $U_3>U_2=U_4$. In the four-band tunneling case, the transmission decreases in $T^+_+$,  $T^-_+$ and  $T^-_-$ channels in comparison with the single barrier case. It does, however, increase for $T^+_-$ when compared to the single barrier case. Transmission is suppressed in the gap region when an interlayer bias is introduced. {This is reflected in the total conductance $G_{\text{tot}}$ in the region of zero conductance}. Our results are relevant for electron confinement in AB bilayer graphene and for the development of graphene-based transistors.


\end{abstract}          

\maketitle

\section{Introduction}
{Numerous theoretical and experimental studies have been conducted on graphene \cite{graphene} since its discovery in 2004.
	This is because of its remarkable features on the one hand and its broad range of potential applications in both fundamental and technological sciences on the other. Unlike ordinary materials, graphene exhibits unusual electronic properties. In fact, it has a very high electronic mobility of 250 000 cm$^2$/Vs, which is 20 times greater than that of GaAS \cite{Novoselov,Pulizzi,Bolotin,Morozov,Gosling}. In addition, in graphene, charge carriers behave like relativistic massless particles and possess a linear energy dispersion at the Dirac points \cite{Katsnelson,Zhang,Neto,Gusynin,Farjadian}. Moreover, graphene has a zero band gap energy \cite{Peng,Cheung,Joaud}, an unconventional quantum Hall effect \cite{Gusynin, Novoselov,Tan,Anirban} and a minimum conductivity \cite{Novoselov,Tworzydlo}. Another remarkable property of graphene is the phenomenon known as Klein tunneling, where electrons can tunnel perfectly through a potential barrier \cite{Katsnelson,Young,Stander,Korol,Correa,Nadi}}. 
Graphene is a promising material for applications such as carbon-based transistors \cite{Lemme, Lin, Cheng, Zhan,Duan,Hussain}, optoelectronic devices \cite{Bonaccorso, Liu, Sun}, and strain sensors \cite{Jeong, Zhao, Tian, Chun} due to its high electronic mobility, high thermal conductivity \cite{Balandin, Geim}, and optical and mechanical properties \cite{Nair, Lee,Ömer}. However, since the graphene conduction and valence bands touch each other, an energy gap must be opened to confine electrons in graphene and enable transistors to switch off \cite{Tudorovskiy,Yung}. One of the techniques to open and control a gap is the application of an external electric field to AB bilayer graphene \cite{McCann,Abergel,Tang}.

AB bilayer graphene results from the stacking of two layers of graphene in accordance with Bernal's method \cite{Bernal}. Similar to monolayer graphene, AB bilayer graphene has several intriguing characteristics. It does, in fact, display a peculiar quantum Hall effect \cite{Schedin,Falko,Alem} that is distinct from the Hall effect seen in monolayer graphene. In contrast to monolayer graphene, electron tunneling in bilayer graphene is characterized by anti-Klein tunneling \cite{Katsnelson,Duppen,Dell}, i.e., a perfect reflection. Its energy spectrum shows four parabolic bands \cite{McCann,Koshino}. Two of them touch each other at zero energy, and two others are separated by an energy equal to the interlayer coupling $\gamma_1=0.4$ eV \cite{McCann,Li}. The ability to open and control a gap makes AB bilayer graphene a major asset for nanoelectronic applications \cite{Li,Zasada,Gold}.

Many studies on the transport properties of bilayer graphene for two-band $(E<\gamma_1)$ and four-band $(E>\gamma_1)$ tunneling have been published recently \cite{Barbier,Van,JELLAL2015149,Hassane,Nadia,Mouhafid,Redouani}.
{Indeed, it has been shown that for a double-barrier structure, the presence of bound states in the well is responsible for the presence of transmission resonances in the gap region\cite{Mouhafid,Redouani}. This is not compatible with the development of graphene-based transistors \cite{Mouhafid,Redouani}. In a recent study, we demonstrated the possibility of completely suppressing transmission in the gap region using a triple barrier system \cite{Hassane}. However, in the case of energies smaller than $\gamma_1$, we have found that a bias must be applied in at least two of the three regions of the triple barrier system, considering large barrier widths to achieve the gap. In this work, we aim to enhance the conditions under which a gap can be obtained in AB bilayer graphene. Hence, we study electron tunneling in AB bilayer graphene through a triple potential barrier in the presence of a magnetic field. Our study revealed the following points. (i): The use of the magnetic field makes it possible to study different configurations of the triple barrier, yielding different results. (ii): Transmission increases for rising barrier ($U_2<U_4$) while it decreases for declining barrier ($U_2>U_4$). (iii) A gap is opened for some configurations by applying a bias only in region 3 (where the magnetic field is present) and this does not require large barrier widths. (iv): For a fixed energy, the triple barrier configurations in the presence of a magnetic field enhanced the transmission with multiple oscillations when barrier widths are increased. (v): For energy $E>\gamma_1$, the transmission is completely suppressed in the gap region and this is reflected in the conductance $G_{\text{tot}}$.}

More precisely, first, we assessed the case where the energy is inferior to $\gamma_1$. It was discovered that in the triple barrier case and for $U_2<U_4$, a resonance appears at normal incidence for energies less than $U_4$, which is not the case for the single barrier \cite{JELLAL2015149}.
The number of resonances is found to increase as the value of $U_4$ increases. Transmission is zero in the same region when $U_2>U_4$, and anti-Klein tunneling increases with $U_2$ but does not change when $U_3$ changes. An interesting number of oscillations have appeared in the triple barrier cases in the energy region greater than $U_4$, particularly when $U_3>U_2=U_4$ and when $U_3<U_2=U_4$. 
These oscillation numbers are higher than those obtained in refs. \cite{Van,JELLAL2015149,Hassane,Nadia,Mouhafid}. When we fix the value of energy at $E=0.39\gamma_1$ and increase the barrier widths, Klein-tunneling decreases and the number of oscillations increases more, even at non normal incidence. 
When $U_2<U_3<U_4$ and $U_3>U_2=U_4$ are used, an interlayer bias opens a gap, whereas it has no effect when $U_2>U_4$, $U3<U_2<U_4$, and $U3<U_2=U_4$ are used. Subsequently, we have considered the case of energy superior to $\gamma_1$. The transmission decreased in the channels $T^+_+$,  $T^-_+$ and  $T^-_-$ compared to the result in a single barrier \cite{JELLAL2015149} while in the $T^+_-$ channel it increased. Transmission is zero in the gap region for a non-zero interlayer bias of $(\delta_3=0.3\gamma_1) $, contrary to the findings in refs. \cite{Mouhafid, Lu, Redouani}. 
{Note that the transmission observed in the gap region in the symmetrical and asymmetrical double barrier cases arises from the bound states present in the well, which is not suitable for graphene-based electronic \cite{Mouhafid}. In the triple-barrier case, and especially in configurations where $U_3$ does not act as a well, there is a complete suppression of transmission in the gap region. This shows the relevance of the triple barrier system, as it provides the on/off switch state for grraphene-based transistors.}


 This work is organized as follows. In Sec. \ref{TTMM}, we present the theoretical model and determine the eigenvectors and eigenvalues of the system. Sec. \ref{Trans} deals with the determination of the transmission probabilities using the continuity conditions and the transfer matrix method, {leading to the corresponding conductance}. In Sec. \ref{RRDD}, we numerically present and discuss the obtained results. Finally, we provide a summary of the results in Sec. \ref{CC}.

\section{Theory and methods}\label{TTMM}
\begin{figure}[H]
        \centering
        \includegraphics[scale=0.25]{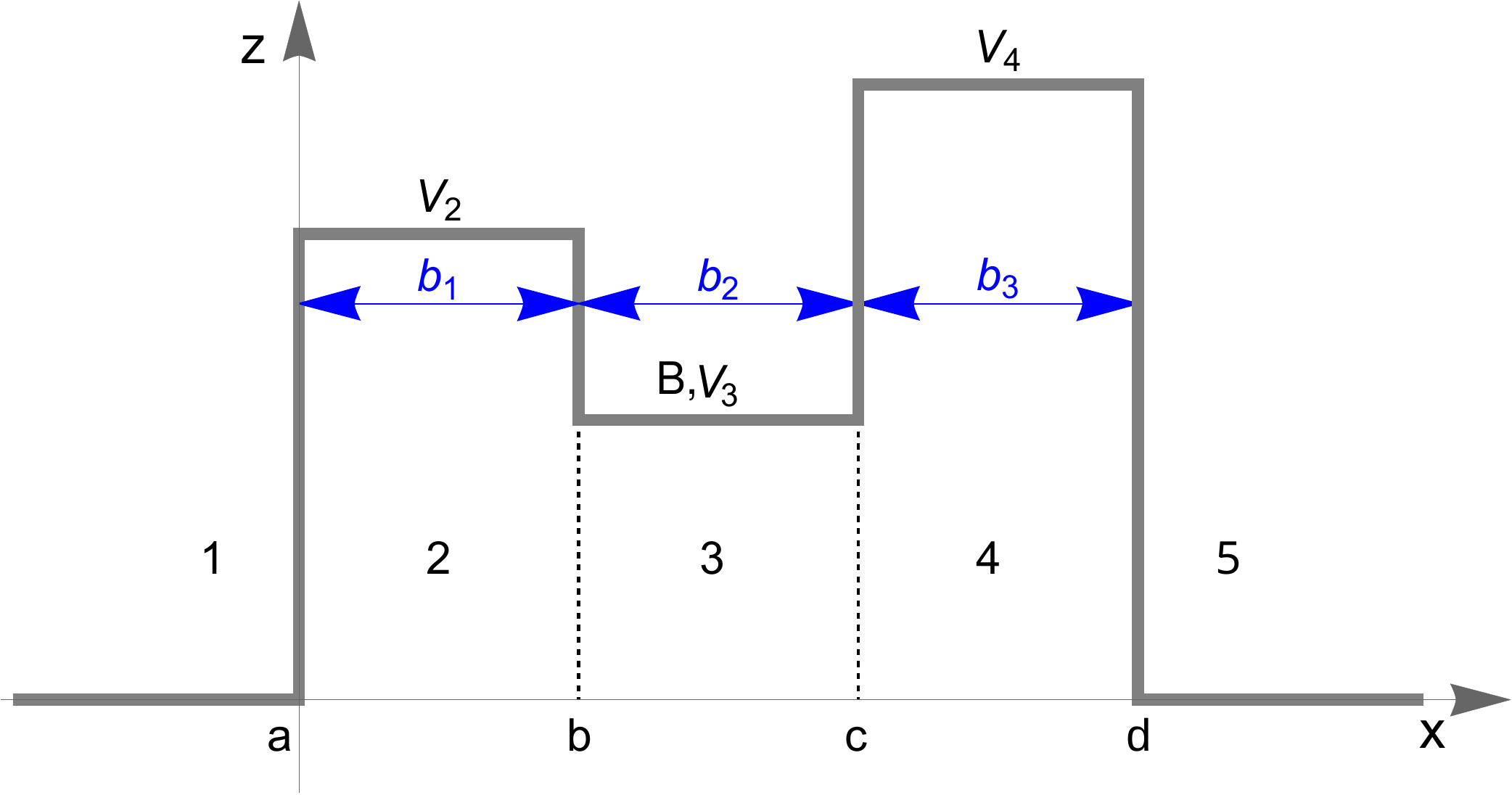}
        \caption{Schematic representation of five regions, including the triple barrier and magnetic field.}
        \label{fig:png2pdf}
\end{figure}

We consider a system of triple electrostatic barriers through which electrons in bilayer graphene can tunnel. As depicted in Fig. \ref{fig:png2pdf}, our system is made up of five regions, each denoted by $  j $. In addition to the electric field, the central region $ (j=3) $ is subjected to a magnetic field. 

The following Hamiltonian  can describe our system \cite{Neto,Van}:
\begin{equation}
        H=\begin{pmatrix}
                V^{+} & v_{F} \pi^{\dagger} & -v_{4} \pi^{\dagger} & v_{3} \pi \\
                v_{F} \pi & V^{+} & \gamma_{1} & -v_{4} \pi^{\dagger} \\
                -v_{4} \pi & \gamma_{1} & V^{-} & v_{F} \pi^{\dagger} \\
                v_{3} \pi^{\dagger} & -v_{4} \pi & v_{F} \pi & V^{-}
        \end{pmatrix}
\label{Eq1}
\end{equation}
where $\pi=p_{x}+i( p_{y}+eA_{y}(x))$ is the in-plan momenta, $v_{F}=10^{6}$ m/s is the Fermi velocity. The potentials $V^{+}$ and  $V^{-}$ on the first and second layers, respectively, are defined by  
\begin{equation}
        V_{j}^{\pm}= \begin{cases}0, & j=1,5 \\ U_{j} + \xi \delta_{j}, & j=2,3,4\end{cases}
        \label{Eq2}
\end{equation}
such that $\xi=+1$ and $\xi=-1$ correspond to the first and second layers. 
The barrier strength is represented by the parameter $U_j$, and the interlayer bias is represented by the parameter $\delta_j$. 
The magnetic field is applied perpendicularly to the graphene layers, and it can be written in terms of the Heaviside step function $\Theta$
\begin{equation}
        B(x, y)=B \Theta\left[\left(b-x\right)\left(c-x\right)\right]
        \label{Eq3}
\end{equation}
where $B$ is a constant. The component of the vector potential $A_y(x)$ in the Landau gauge is defined as  
\begin{equation}
        A_{y}(x)=\frac{\hbar}{e l_{B}^{2}} \begin{cases}
                b, & \text{if} \quad x<b \\
                x, & \text{if}  \quad b<x<c \\
                c, & \text{if} \quad x>c 
    \end{cases}
        \label{Eq4}
\end{equation}
where $l_{B}=\sqrt{\hbar/eB}$ is the magnetic length. The parameters $v_{3}$ and $v_{4}$ have been shown to have no effect on the band structure at high energy or on the transmission at low energy \cite{McCann,Van,Hosein}. Then, we neglect them and write the Hamiltonian (\ref{Eq1}) as
\begin{equation}
        H_{j}= \begin{pmatrix}
                U_{j}+\delta_{j} & v_{F} \pi^{\dagger} &0  &0 \\
                v_{F} \pi &     U_{j}+\delta_{j} & \gamma_{1} & 0  \\
                0 & \gamma_{1} &        U_{j}-\delta_{j} & v_{F} \pi^{\dagger} \\
                0 & 0  &        v_{F} \pi &     U_{j}-\delta_{j}
        \end{pmatrix}
        \label{Eq5}
\end{equation} 
in the basis of four component spinor 
$ 
        \psi^{j}(x, y)=\left[\psi_{A_{1}}^{j}, \psi_{B_{1}}^{j}, \psi_{A_{2}}^{j}, \psi_{B_{2}}^{j}\right]^{\dagger}
 $
where the symbol $\dagger$ denotes the transpose row vector. As a consequence of the conservation of momentum $k_y$ along the $ y $-direction, the spinor  can be written as
\begin{equation}
        \psi^{j}(x, y)=e^{i k_{y}y}\left[\phi_{A_{1}}^{j}, \phi_{B_{1}}^{j}, \phi_{A_{2}}^{j}, \phi_{B_{2}}^{j}\right]^{\dagger}
        \label{Eq7}
\end{equation}

       
       In order to determine the eigenvalues and the eigenvectors of each region, we use    $  H_{j} \psi_{j}=E_{j} \Psi_{j} $ to obtain four coupled differential equations
\begin{subequations}
        \begin{align}
                -i \sqrt{2} \vartheta_{0} a \phi_{B_{1}}^{j}&=(\varepsilon_{j}-\delta_{j})\phi_{A_{1}}^{j}\label{Eq9a}
                \\
                i \sqrt{2} \vartheta_{0} a^{\dagger} \phi_{A_{1}}^{j}&=(\varepsilon_{j}-\delta_{j})\phi_{B_{1}}^{j}-\gamma_{1}\phi_{A_{2}}^{j} \label{Eq9b}
                \\
                -i \sqrt{2} \vartheta_{0} a \phi_{B_{2}}^{j}&=(\varepsilon_{j}+\delta_{j})\phi_{A_{2}}^{j}-\gamma_{1}\phi_{B_{1}}^{j}  \label{Eq9c} 
                \\
                i \sqrt{2} \vartheta_{0} a^{ \dagger} \phi_{A_{2}}^{j}&=(\varepsilon_{j}+\delta_{j})\phi_{B_{2}}^{j}
                \label{Eq9d}
        \end{align}
\end{subequations}
where we have set $\varepsilon_{j}=E_{j}-U_{j}$, $\vartheta_{0}=\frac{ \hbar v_{F}}{lb}$ is the energy scale, $a$ and $a^{\dagger}$ are respectively the annihilation and creation operators 
\begin{subequations}
        \begin{align}
               & a=\frac{l_{B}}{\sqrt{2}}\left(\partial_{x}+k_{y} +\frac{eA_{y}(x)}{\hbar}\right)
                        \label{Eq10a}\\
                &a^{\dagger}=\frac{l_{B}}{\sqrt{2}}\left(-\partial_{x}+k_{y} +\frac{eA_{y}(x)}{\hbar}\right)
                \label{Eq10b}
        \end{align}
\end{subequations}
which satisfy the commutation relation $\left[a, a^{\dagger}\right]=\mathbb{I}$. We eliminate the unknowns from (\ref{Eq9a}-\ref{Eq9d}) step by step and obtain for $\phi_{B_{1}}$ 
\begin{widetext}
\begin{equation}
        \left[ 2\vartheta_{0}^{2}aa^{\dagger}-(\varepsilon_{j}+\delta_{j})^{2}\right]\left[ 2\vartheta_{0}^{2}a^{\dagger}a-(\varepsilon_{j}-\delta_{j})^{2}\right] \phi_{B_{1}}^{j}  =\gamma_{1}^{2}(\varepsilon_{j}^{2}-\delta_{j}^{2}) \phi_{B_{1}}^{j}
        \label{Eq11}
\end{equation}
\end{widetext}

\subsection{Eigenvalues and eigenvectors in  region 3}
In region 3, where the vector potential is $A_{y}(x)=\frac{\hbar x}{e l_{B}^{2}}$, we introduce the variable $z=\sqrt{2}(\frac{x}{l_{B}}+k_{y}l_{B})$ and  solve  (\ref{Eq11}) to obtain $\phi_{B_{1}}$. 
%
Then, we substitute the result into  (\ref{Eq9a}-\ref{Eq9d}) to determine the rest of the spinor components. For more simplicity, we write the solution in matrix form
\begin{equation}
        \psi^3(x, y)=G_3 M_3(x) C_3
        \label{Eq12}
\end{equation}
where $G_{3}=\mathbb{I}_{4}$ is $4\times4$ identity matrix,  $C_{3}=\left( c_{+},c_{-},d_{+},d_{-}\right) ^{\dagger}$ is a constant, and the matrix $M_{3}(x)$ is given by
\begin{widetext}
\begin{equation}
        M_{3}(x)=
        \begin{pmatrix}
                \eta_{-} \lambda_{+} \chi_{+,-1}^{+} & \eta_{-}^* \lambda_{+} \chi_{-,-1}^{+} & \eta_{-} \lambda_{-} \chi_{+,-1}^{-} & \eta_{-}^* \lambda_{-} \chi_{-,-1}^{-} \\
                \chi_{+,0}^{+} & \chi_{-,0}^{+} & \chi_{+,0}^{-} & \chi_{-,0}^{-} \\
                \zeta^{+} \chi_{+,0}^{+} & \zeta^{+} \chi_{-,0}^{+} & \zeta^{-} \chi_{+,0}^{-} & \zeta^{-} \chi_{-,0}^{-}  \\
                \eta_{+}^* \zeta^{+} \chi_{+,1}^{+} & \eta_{+} \zeta^{+} \chi_{-,1}^{+}  & \eta_{+}^* \zeta^{-} \chi_{+,1}^{-} & \eta_{+} \zeta^{-} \chi_{-,1}^{-} 
       \end{pmatrix}
\label{Eq13}
\end{equation}
\end{widetext}
where $\chi^{\tau}_{\pm, l}= D\left[\lambda_{\tau}\pm l,\pm z \right]$ are the parabolic cylindrical function with argument $z$ and $ l=-1, 0, 1 $. We have set the quantities
\begin{align}
   &     \lambda_{\tau}=-\frac{1}{2}+\frac{\varepsilon_{3}^{2}+\delta_{3}^{2}}{2\vartheta_{0}^{2}}+\tau\frac{\sqrt{(\vartheta_{0}^{2}-2\varepsilon_{3}\delta_{3})^{2}+\gamma_{1}^{2}(\varepsilon_{3}^{2}-\delta_{3}^{2})}}{2\vartheta_{0}^{2}}
        \label{Eq14}
\\
&
        \eta_{\pm}=\frac{-i\sqrt{2}\vartheta_{0}}{\varepsilon_{3}\pm\delta_{3}}
        \label{Eq15}
\\
 &        \zeta^{\pm}=\frac{-2\vartheta_{0}^{2}\lambda_{\pm}+(\varepsilon_{3}-\delta_{3})^{2}}{\gamma_{1}(\varepsilon_{3}-\delta_{3})}
                \label{Eq16}
\end{align}
We solve  (\ref{Eq14}) to obtain the energy spectrum in this region. It is given by
\begin{widetext}
        \begin{equation}
                        \varepsilon^{\tau}_{\pm,3}= \pm \frac{1}{\sqrt{6}}\left[\mu^{\frac{1}{3}}+\nu \mu^{\frac{-1}{3}}+2 A\right]^{\frac{1}{2}}\\
                +\tau\frac{1}{\sqrt{6}}\left[-6 B \sqrt{6}\left(\mu^{\frac{1}{3}}+\nu \mu^{\frac{-1}{3}}+2 A\right)^{\frac{-1}{2}}-\left(\mu^{\frac{1}{3}}+\nu \mu^{\frac{-1}{3}}-4 A\right)\right]^{\frac{1}{2}}                      \label{Eq17}
                \end{equation}
\end{widetext}
where we have defined the parameters
        \begin{align}
&       \mu=-A^3+27 B^2+9 A C+
        \sqrt{\left(-A^3+27 B^2+9 A C\right)^2-\nu^3}
        \label{Eq18}
\\
&
\nu=(A^{2}+3C)
        \label{Eq19}
\\
&       
A=\delta_3^2+(2 n+1) \vartheta_0^2+\frac{\gamma_1^2}{2}
\label{Eq20}
\\ 
&
        B=\vartheta_0^2 \delta_3 
        \label{Eq21}
\\
&       C=\left((2 n+1) \vartheta_0^2-\delta_3^2\right)^2-\vartheta_0^4+\gamma_1^2 \delta_3^2 
 \label{Eq22}
\end{align}
and $n=\lambda_{\tau}$ is an integer number.

\subsection{Eigenvalues and eigenvectors in  regions 1, 2, 4, 5 }

In regions $ j=1, 2, 4, 5 $, the vector potential is constant and set to be $A_{y}(x)=\frac{ \hbar }{e l_{B}^{2}}d_{j}$ where 
\begin{equation}
d_{j}=\left\{\begin{array}{lll}
        b, & \text { if } & x<b \\
        c, & \text { if } & x>c
\end{array}\right.
\label{Eq23}
\end{equation}
By solving  (\ref{Eq11}) for $\phi_{B_{1}}$, and substituting the result into  (\ref{Eq9a}-\ref{Eq9d}), we obtain a general solution  in the matrix form 
\begin{equation}
        \psi^j(x, y)=G_j M_j(x) C_j e^{i k_y y}
        \label{Eq24}
\end{equation}
where  $G_{j}$ and $M_{j}(x)$ are igiven by 
\begin{align}
    &    G_j=\begin{pmatrix}
                f_{-}^{+} & -f_{+}^{+} & f_{-}^{-} & -f_{+}^{-} \\
                1 & 1 & 1 & 1 \\
                h^{+} & h^{+} & h^{-} & h^{-} \\
                h^{+} g_{+}^{+} & -h^{+} g_{-}^{+} & h^{-} g_{+}^{-} & -h^{-} g_{-}^{-}
        \end{pmatrix}
\label{Eq25}
\\
&
        M_j(x)=\begin{pmatrix}
                e^{i k_j^{+} x} & 0 & 0 & 0 \\
                0 & e^{-i k_j^{+} x} & 0 & 0 \\
                0 & 0 & e^{i k_j^{-} x} & 0 \\
                0 & 0 & 0 & e^{-i k_j^{-} x}
        \end{pmatrix}
\label{Eq26}
\end{align}
with the parameters
\begin{align}
&
        f_{\pm}^\tau=\hbar v_F \frac{k_j^\tau \pm i\left(k_y+\frac{d_j}{l_B^2}\right)}{\varepsilon_j-\delta_j}
        \label{Eq27}
\\
&
        h^\tau=\frac{\left(\varepsilon_j-\delta_j\right)^2-\left(\hbar v_F\right)^2\left[\left(k_j^\tau\right)^2+\left(k_y+\frac{d_j}{l_B^2}\right)^2\right]}{\left(\varepsilon_j-\delta_j\right) \gamma_1}
                \label{Eq28}
\\
&
        g_{\pm}^\tau=\hbar v_F \frac{k_j^\tau \pm i\left(k_y+\frac{d_j}{l_B^2}\right)}{\varepsilon_j+\delta_j}
                \label{Eq29}
\end{align}
The wave vector $k_{j}$ along the $ x $-direction is expressed as 
\begin{equation}
        k_j^\tau=\sqrt{\frac{\varepsilon_j^2+\delta_j^2+\tau \sqrt{4 \varepsilon_j^2 \delta_j^2+\gamma_1^2\left(\varepsilon_j^2-\delta_j^2\right)}}{\left(\hbar v_F\right)^2}-\left(k_y+\frac{d_j}{l_B^2}\right)^2}
        \label{Eq30}
\end{equation}
giving rise to the eigenenergies
\begin{equation}
        \varepsilon_{\pm, j}^\tau=\pm \sqrt{\delta_j^2+\left(\hbar v_F k\right)^2+\frac{\gamma_1^2}{2}+\tau \sqrt{\left(\hbar v_F k\right)^2\left(4 \delta_j^2+\gamma_1^2\right)+\frac{\gamma_1^4}{4}}}
        \label{Eq31}
\end{equation}
where $k=\left[\left(k_j^\tau\right)^2+k_y^2\right]^{\frac{1}{2}}$ is the wave vector. It is worth noting that in the incident and transmission regions $ j = 1, 5 $, solutions are obtained by requiring $U_{j}=\delta_{j}=0$.

\section{Transport properties}\label{Trans}

We will calculate the transmission probability corresponding to the present system. We do this by imposing continuity conditions at each interface of the triple barrier structure. Thereafter, we can use the transfer matrix method to make a connection between the coefficients of the incident region and those of the transmitted one. These coefficients are given by
\begin{equation}
C_{\text{1}}^{\tau}=\begin{pmatrix}
        \delta_{\tau, 1} \\
        r_{+}^{\tau} \\
        \delta_{\tau,-1} \\
        r_{-}^{\tau}
\end{pmatrix},\quad
C_{\text{5}}^{\tau}=\begin{pmatrix}
        t_{+}^{\tau} \\
        0 \\
        t_{-}^{\tau} \\
        0
\end{pmatrix}
\label{Eq32}
\end{equation}
where $\delta_{\tau,\pm1}$ is the Kronecker delta symbol. The continuity at interfaces $x=a, b, c, d$ gives rise to
\begin{align}
&
G_{1} M_{1}(a) C_{1}=G_{2} M_{2}(a) C_{2} 
\label{Eq33}\\
&
G_{2} M_{2}(b) C_{2}=G_{3} M_{3}(b) C_{3} 
\label{E34}\\
&
G_{3} M_{3}(c) C_{3}=G_{4} M_{4}(c) C_{4}
\label{E35} \\
&
G_{4} M_{4}(d) C_{5}=G_{5} M_{5}(d) C_{5}
\label{E36}.
\end{align}
We can now  connect the coefficients $C_{\text{1}}^{\tau}$ to $C_{\text{5}}^{\tau}$
\begin{equation}
C_{\text{1}}^{\tau}=N C_{\text{5}}^{\tau}
\label{E37}
\end{equation}
where  the matrix transfer $N$ takes the form
\begin{equation}
N=\prod_{j=\text{1}}^{\text{4}} M_{j}^{-1}(x_{j}) G_{j}^{-1} G_{j+1} M_{j+1}(x_{j})
\label{E38}
\end{equation}
After manipulation, we can write  (\ref{E37}) as
\begin{equation}
\begin{pmatrix}
        t_{+}^{\tau} \\
        r_{+}^{\tau} \\
        t_{-}^{\tau} \\
        r_{-}^{\tau} 
\end{pmatrix}=\begin{pmatrix} 
        N_{11} & 0 & N_{13} & 0 \\
        N_{21} & -1 & N_{23} & 0 \\
        N_{31} & 0 & N_{33} & 0 \\
        N_{41} & 0 & N_{43} & -1
\end{pmatrix}^{-1} \begin{pmatrix} 
        \delta_{\tau, 1} \\
        0 \\
        \delta_{\tau,-1} \\
        0
\end{pmatrix}
\label{E39}
\end{equation}
where $N_{ij}$ are the matrix elements of $N$ in (\ref{E38}). Then, we can easily derive the transmission coefficients from  (\ref{E39}). They are given by 
\begin{align}
&t_{+}^{\tau}=\frac{N_{33} \delta_{\tau, 1}-N_{13} \delta_{\tau,-1}}{N_{11} N_{33}-N_{13} N_{31}}
\label{E40}
\\
& t_{-}^{\tau}=\frac{N_{11} \delta_{\tau,-1}-N_{31} \delta_{\tau, 1}}{N_{11} N_{33}-N_{13} N_{31}} 
\label{E41}
\end{align}

At this stage, we have to introduce the current density to find all transmission channels, which is
\begin{equation}
\textbf{j}=v_{F}{\Psi}^{\dagger}\vec{\alpha}\Psi
\label{E42}
\end{equation}
where $\vec{\alpha}$ is a $4\times4$ matrix with two Pauli matrices $\sigma_{x}$ on the diagonal, and the rest is  zero. Then, using \eqref{E42}, we can calculate 
the incident $\textbf{j}_{\text{inc}}$ and transmitted $\textbf{j}_{\text{tra}}$ current densities. As a result, we obtain the four transmission  channels
\begin{align}
&     T_{\pm}^{\tau}=\frac{\left|\textbf{j}_{\text{tra}}\right|}{\left|\textbf{j}_{\text{inc}}\right|}=\frac{k_{5}^{\pm}}{k_{1}^{\tau}}\left|t_{\pm}^{\tau}\right|^{2}
\label{E43}
\end{align} 

{Afterward, we will use the  Landauer–Büttiker formula to calculate the conductance based on the previously determined transmissions. This is
\begin{equation}
\mathrm{G}(E)=G_0 \frac{L_y}{2 \pi} \int_{-\infty}^{+\infty} \mathrm{d} k_y \sum_{\tau, n= \pm} T_n^\tau\left(E, k_y\right)
\end{equation}
where $G_0=$ $4 \frac{e^2}{h}$, the factor 4 arises from the valley and spin degeneracy in graphene. $L_y$ is the width of the sample in the $y$-direction.}

\section{Results and discussions}\label{RRDD}
In the following, we will compute and discuss our numerical results. To begin with, we will look at the case of energy less than the interlayer coupling $\gamma_{1}$ (two-band tunneling). Here, only one mode of propagation is possible, so we have just one transmission channel. Next, we will consider energy greater than the interlayer coupling $\gamma_{1}$ (four-band tunneling). In this case, two propagation modes are available, which result in four transmission channels.

\subsection{Two-band Tunneling}

\begin{figure}[H]
	\centering 
	\includegraphics[scale=0.68]{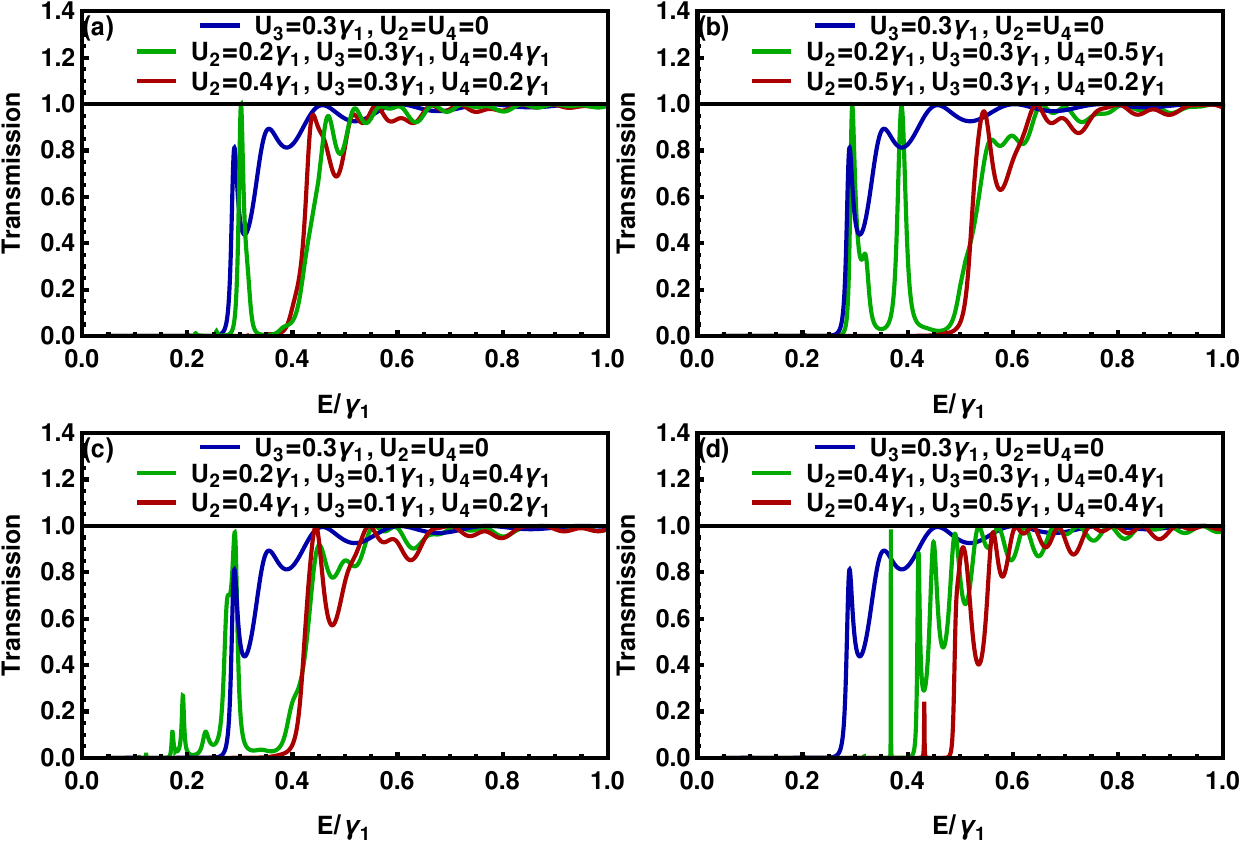}
	\caption{(Color online): Transmission  as a function of energy $ E $, at normal incidence ($k_y=0$), for barrier widths $b_{1}=b_{2}=b_{3}= 25$ nm, $l_B=13.5$ nm, and $\delta_2=\delta_ 3=\delta_ 4=0$. (a): $U_2<U_3<U_4$ (green line) and $U_2>U_3>U_4$ (red line). (b): As in (a) except that $U_4= 0.5 \gamma_1$ (green line) and $U_2= 0.5 \gamma_1$ (red line). (c): $U_3<U_2<U_4$ (green line) and  $U_2>U_4>U_3$ (red line). (d): $U_3<U_2=U_4$ (green line) and  $U_3>U_2=U_4$ (red line). The blue line corresponds to the result obtained in  \cite{JELLAL2015149} for a single barrier. 
	}
	\label{Transmission}
\end{figure}

The transmission probability is shown in Fig. \ref{Transmission} as a function of incident energy $E$ at normal incidence $(k_y=0)$ for barrier widths of $b_1=b_2=b_3=25$ nm. The magnetic length is set to $ l_B = 13.5 $ nm, as in the single barrier case discussed in \cite{JELLAL2015149}, and the interlayer bias is set to $\delta_2=\delta_3=\delta_4=0$. We plot the transmission for two configurations of the triple barrier system in Fig. \ref{Transmission}(a), and compare it to the result found in \cite{JELLAL2015149} (blue line). In contrast to the single barrier case (blue line), a transmission resonance occurs for energy lower than $U_4$ when $U_2<U_3<U_4$ (green line). 
In the same energy range, however, transmission is zero (anti-Klein tunneling \cite{Katsnelson,Duppen}) when $U_2>U_3>U_4$ (red line). The transmission oscillates faster in the triple barrier cases (green and red lines) than in the single barrier (blue line \cite{JELLAL2015149}) and double barrier for zero magnetic field \cite{Mouhafid} cases for energies greater than $ U_4 $. {Remember that anti-Klein tunneling in bilayer graphene is caused by the fact that the barrier cloaks the confined states, making them invisible to transmission \cite{GU}. In the case of the triple barrier, anti-Klein tunneling depends only on the value of the first barrier, i.e., the value of $U_2$. In fact, in the case where $U_2<U_3<U_4$, the value of $U_2$ is small, and then anti-Klein tunneling is limited to $E=0.2\gamma_1$, whereas for the opposite configuration, due to the large value of $U_2=0.4 \gamma_1$, anti-Klein tunneling extends to $E=0.4\gamma_1$.  In addition, the resonance observed at $E=0.3\gamma_1$ for  $U_2<U_3<U_4$ results from hole states inside the barriers $U_3$ and $U_4$ through which the electrons can tunnel. In the case where $U_2>U_3>U_4$, because of the magnitude of anti-Klein tunneling, no such resonance is observed.}
Fig. \ref{Transmission}(b) depicts the transmission with the same parameters as in Fig. \ref{Transmission}(a), but with $U_4$ (green line) and $U_2$ (red line) increased. The number of resonances increases as $U_4$ increases for configuration  $U_2<U_3<U_4$ (green line). {Indeed, increasing the barrier height ($U_4$) increases the number of hole states through which electrons can tunnel, thus increasing the number of resonances.} In the opposite configuration, we see that anti-Klein tunneling increases as $U_2$ increases. {This confirms the dependence of anti-klein tunneling on the first barrier ($U_2$). } In Fig. \ref{Transmission}(c), we reduce $U_3$ so that $U_3<U_2<U_4$ (green line) and $U_2>U_4>U_3$ (red line) are obtained. Transmission occurs for energies less than $U_2$ for $U_3<U_2<U_4$ (green line), but not for $U_3$ (blue line see \cite{JELLAL2015149}). {Note that when the value of $U_3$ is reduced, it acts like a well between $U_2$ and $U_4$. Then, the bound states present in the well are responsible for the non-zero transmission observed in the region preceding $E=U_2$ \cite{Barbier,Mouhafid}.} {For $U_2>U_4>U_3$ (red line), due to the large magnitude of anti-Klein tunneling, decreasing $U_3$ has no significant effect on the transmission, which remains nearly identical to that shown in Fig. \ref{Transmission}(a)}. We conclude that when $U_2>U_4$, the transmission decreases regardless of whether $U_3$ is large or small. 
The transmission for $U_3<U_2=U_4$ (green line) and $U_3>U_2=U_4$ (red line) is shown in Fig. \ref{Transmission}(d). In both cases, the number of transmission oscillations increases significantly when compared to the single barrier's result \cite{JELLAL2015149}. However, when $U_3>U_2=U_4$, it is less than when $U_3<U_2=U_4$. {This difference arises from the bound states available between $U_2$ and $U_4$ when $U_2=U_4<U_3$ and which are absent in the opposite case}. It is also worth noting that such a large number of oscillations is not observed in  \cite{Van,Hassane,Nadia, Mouhafid}.

\begin{figure}[H]
	\centering
	\includegraphics[scale=0.68]{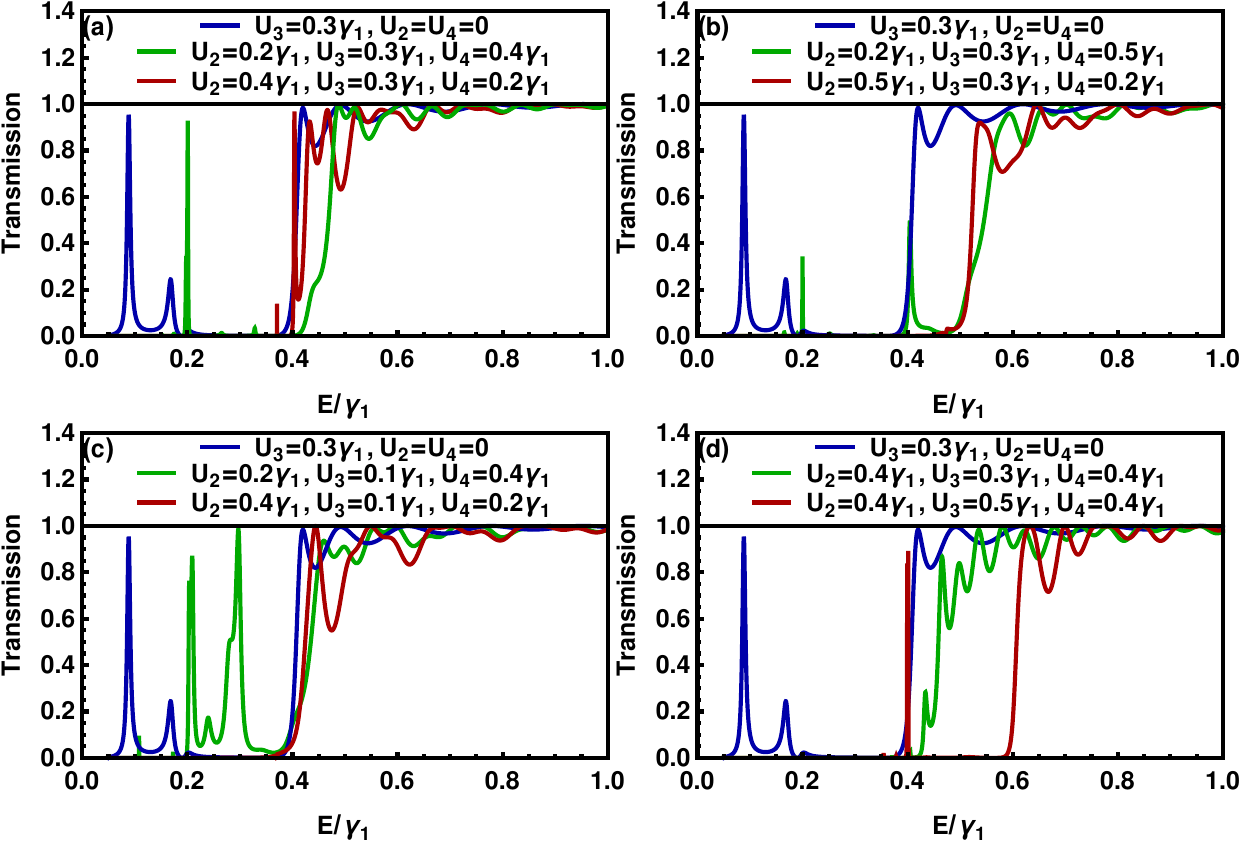}
	\caption{(Color online): The same parameters as in Fig. \ref{Transmission} but now for $\delta_3=0.1\gamma_1$ and $\delta_2=\delta_4=0$.
	}
	\label{Trans&with&biais}
\end{figure}

To investigate the interlayer bias effect, we plot the transmission as a function of incident energy $E$ in Fig. \ref{Trans&with&biais}, with the same parameters as in Fig. \ref{Transmission} and $\delta_3=0.1\gamma_1$ is set. The presence of the interlayer bias in Fig. \ref{Trans&with&biais}(a,b) opens a gap for the triple barrier system when $U_2<U_3<U_4$ (green line), as it does for the single barrier system (blue line \cite{JELLAL2015149}). In contrast, there is no gap when $U_2>U_3>U_4$ (red line), and the transmission behaves similarly to the result in Fig. \ref{Transmission} with a minor difference. {Applying a bias only in region 3 is not enough to obtain a gap in this case, as in \cite{Hassane}}. The interlayer bias has no meaningful effect on transmission in Fig. \ref{Trans&with&biais}(c) when $U_3<U_2<U_4$ or when $U_2>U_4>U_3$, and no gap is found. {In the first case, the presence of bound states in the well is responsible for the resonances found instead of a gap \cite{Mouhafid}. In the second case, as when $U_2>U_3>U_4$, the applied bias is not sufficient to open a gap, then the anti-Klein tunneling remains.} However, in Fig. \ref{Trans&with&biais}(d), there is a gap when $U_3>U_2=U_4$, but not when $U_3<U_2=U_4$. {This difference arises from the absence of bound states when $U_3>U_2=U_4$ while they are present when $U_3<U_2=U_4$.} Note that, in the triple barrier case, the transmission does not behave like the results in \cite{JELLAL2015149,Hassane,Nadia,Mouhafid} {when a gap is opened}. For instance, the transmission is not zero in the region preceding the gap in \cite{JELLAL2015149,Hassane,Nadia}, while it is zero in the triple barrier case.
Let us also note that in the case of a triple barrier with zero magnetic field \cite{Hassane}, it is necessary to apply an interlayer bias in at least two regions to obtain a gap, whereas in this case $\delta_3$ can open a gap on its own {for some configurations}. {Consequently, a bias can be applied in region 2 ($\delta_2\neq 0$) or region 4 ($\delta_4\neq0)$ to obtain a gap in configurations where it is not obtained by $\delta_3=0.1 \gamma_1$}.

\begin{figure}[H]
	\centering
	\includegraphics[scale=0.5]{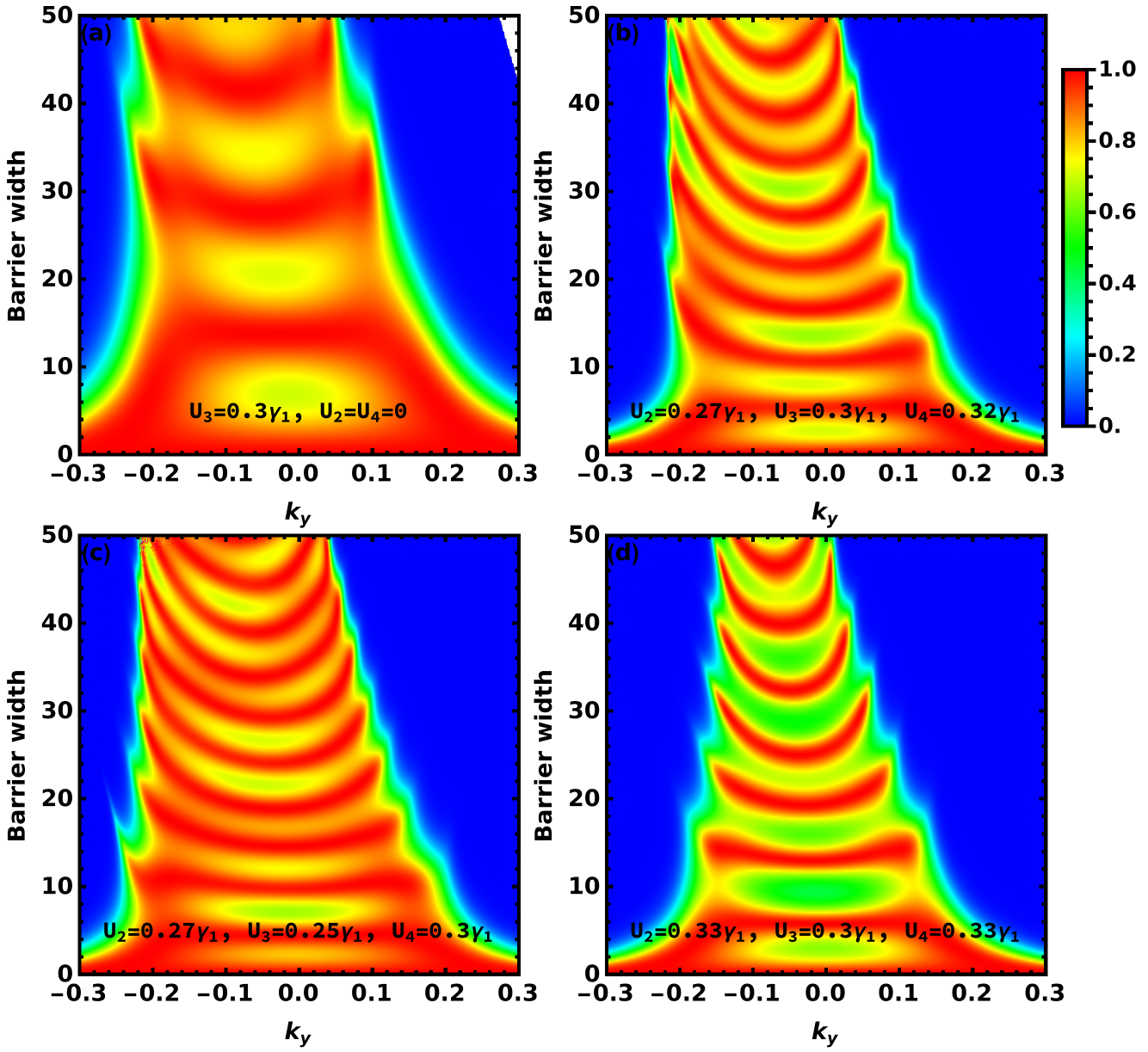}
	\caption{(Color online): Density plot of transmission as a function of the barrier width ($b_1=b_2=b_3$) and the transverse wave vector $k_y$ for $l_B=18.5$ nm and $E=0.39\gamma_1$. (a): Single barrier case, (b): $U_2<U_3<U_4$, (c): $U_3<U_2<U_4$ and (d): $U_3<U_2=U_4$. The interlayer bias are set at $\delta_1=\delta_2=\delta_3=0.$
	}
	\label{DensityBarrierWidthvsky.pdf}
\end{figure}

In Fig. \ref{DensityBarrierWidthvsky.pdf} we show the transmission as a function of barrier width ($b_1=b_2=b_3$) and the transverse wave vector $k_y$. The magnetic length is set at $l_B=18.5$ nm, the interlayer bias at $\delta_1=\delta_2=\delta_3=0$, and the energy at $E=0.39\gamma_1$. When we compare Fig. \ref{DensityBarrierWidthvsky.pdf}(a) (single barrier case \cite{JELLAL2015149}) to Fig. \ref{DensityBarrierWidthvsky.pdf}(b) ($U_2<U_3<U_4$), we see that the number of oscillations increases significantly in the triple barrier case as the barrier widths increase. In Fig. \ref{DensityBarrierWidthvsky.pdf}(c) additional transmission resonances appear where $U_3<U_2<U_4$, resulting in more oscillations than in Fig. \ref{DensityBarrierWidthvsky.pdf}(b). The number of resonances, on the other hand, decreases when $U_3<U_2=U_4$, as shown in Fig. \ref{DensityBarrierWidthvsky.pdf}(d). As a result, the oscillations are less frequent than in Fig. \ref{DensityBarrierWidthvsky.pdf}(b). However, it is important to note that the number of oscillations is greater than that of the single barrier \cite{JELLAL2015149} and double-barrier \cite{Mouhafid} cases. Klein tunneling is found to be reduced in the triple barrier cases when compared to the single barrier  \cite{JELLAL2015149}. 
In fact, at $\lvert k_y\rvert=0.3$ nm${^{-1}}$, anti-Klein tunneling begins at $b_1=b_2=b_3=3$ nm in triple barrier cases, and at 8 nm in single barrier case. In addition, for barrier widths greater than 20 nm, the transmission narrows more in the triple barrier cases than it does in the single barrier case. As a consequence, Klein tunneling decreases while anti-Klein tunneling increases.

{It is important to note that the magnetic field plays a crucial role in this work. Indeed, it enables us to investigate different configurations, each yielding unique outcomes. For example, configurations of the types $U_2<U_3<U_4$ and $U_4>U_3>U_2$ provide distinct results. If the magnetic field were zero, they would have the same results, and the study would be restricted. Furthermore, it is noteworthy that, thanks to the presence of the magnetic field, there are configurations where a gap is opened by applying an interlayer bias only in the central region (region 3). In contrast, in the absence of a magnetic field \cite{Hassane} it is found that an interlayer bias must be applied in at least two regions, considering large barrier widths to open a gap. On the other hand, in four-band tunneling, we will study only one configuration of the triple barrier, and asymmetry is expected in all channels of transmission due to the presence of the magnetic field in the central region. }

\subsection{Four-band Tunneling}

\begin{figure}[H]
	\centering
	\includegraphics[scale=0.46]{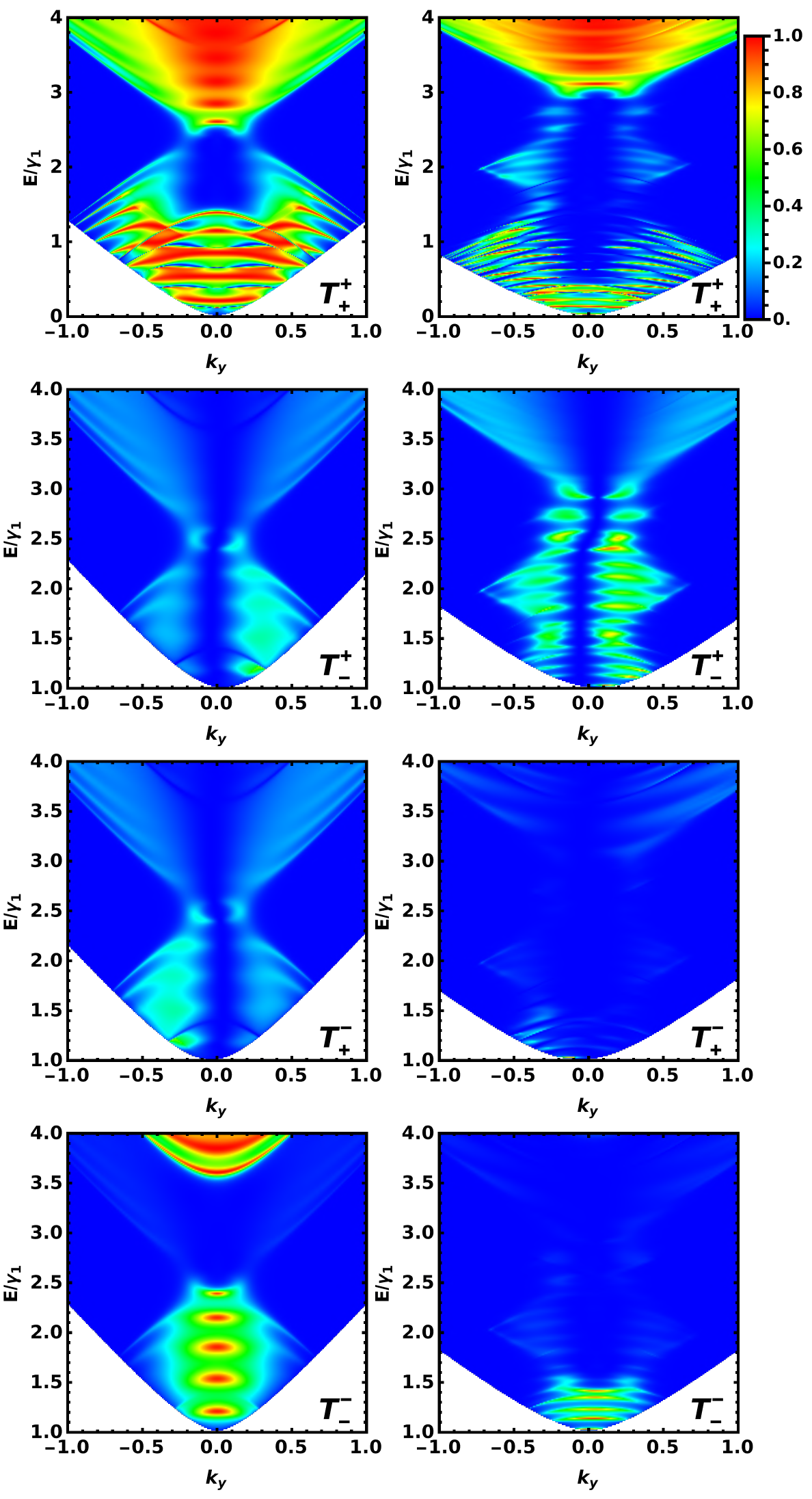}
	\caption{(Color online): Density plot of transmission as a function of incident  energy $E$ and the transverse wave vector $k_y$. (Left): Single barrier case with $U_3=2.5\gamma_1$. (Right): Triple barrier case with $U_2=1.5\gamma_1$, $U_3=2.5\gamma_1$ and $U_4=3\gamma_1$. The barrier widths are set at $b_1=b_2=b_3=15$ nm, $l_B=13.5$ nm and $\delta_2=\delta_3=\delta_4=0.$}
	\label{DensityPlot.pdf}
\end{figure}

We plot the transmission as a function of incident energy $ E $ and the transverse wave vector $k_y$ in Fig. \ref{DensityPlot.pdf} for the single barrier case \cite{JELLAL2015149} with $U_3=2.5\gamma_1$ (left) and the triple barrier case with $U_2<U_3<U_4$ (right). The barrier widths are set to $b_1=b_2=b_3=15$ nm, the magnetic length is set to $l_B=13.5$ nm, and the interlayer bias is set to $\delta_2=\delta_3=\delta_4=0$. 
 The cloak effect \cite{Duppen,Van} occurs in the $T^+_+$ channel at normal incidence in the energy region $U_3-\gamma_1<E<U_3$ ({where the modes $k^{+}$ outside and $k^{-}$ inside the barrier are decoupled}) in the single barrier case, and for a wide range of energy $U_3-2\gamma_1<E<U_3+0.5\gamma_1$ ({where the modes $k^{+}$ outside and $k^{-}$ inside the barrier are decoupled}) in the triple barrier case. {However, at non-normal incidence, the two modes $k^{+}$ outside and $k^{-}$ inside the barrier are coupled, and then transmission occurs. For energies smaller than $U_3-\gamma_1$, there are propagating states $k^{+}$ resulting in non-zero transmission in the single barrier case, whereas for the triple barrier case they are available only for $U_2-\gamma_1$}.  As a result, the transmission resonances decrease for energies less than $U_3$ when compared to the single barrier case. There are, however, more thin resonances than those obtained in \cite{Hassane,Mouhafid,Lu}.
 {Furthermore, the presence of the magnetic field in the triple barrier makes the transmission to be more pronounced for $k_y<0$ in the region $U_3-2\gamma_1<E<U_3+0.5\gamma_1$, while the transmission exhibits symmetrical behavior with respect to normal incidence ($k_y=0$) in  \cite{Van,JELLAL2015149,Hassane,Nadia,Mouhafid}}.
 The transmission probability increases in the $T^+_-$ channel compared to the single barrier case and approaches unity near $E=U_3$ in the case of the triple barrier (right). {This is because the cloak effect appears only at normal incidence.} In addition, the transmission oscillates, unlike \cite{Van,JELLAL2015149,Mouhafid}. {In the $T^-_+$ channel and for the triple barrier case, the cloak effect appears even at non-normal incidence and has large size in contrast to $T^+_-$ channel. As a result, the transmission probability is close to zero, as found in \cite{Hassane}}. In contrast, in the case of a single barrier \cite{JELLAL2015149}, it is different from zero. The transmission in the $T^-_-$ channel decreases from $E=U_3$ in the single barrier case to $E=U_3-\gamma_1$ in the triple barrier case. {Indeed, in $T^{-}_{-}$ channel there are no propagating states $k^-$ in the barriers  in the region upwards of $E=U_3-\gamma_1$, then the cloaking is larger \cite{Van}.} Therefore, the number of resonances diminishes in the triple barrier case, but the transmission probability still remains higher than in \cite{Nadia,Mouhafid}. Resonances appear in the case of a single barrier for energies greater than $U_3+\gamma_1$, whereas the transmission is zero in the case of a triple barrier {due to the cloaking}. {Note that, in the triple barrier in the absence of magnetic field and interlayer bias \cite{Hassane}, in all channels $T^{+}_{+}$, $T^{+}_{-}$,$T^{-}_{+}$ and $T^{-}_{-}$ the transmission is found to be symmetric with respect to normal incidence. In this case, the presence of the magnetic field in region 3 of the triple barrier system breaks the symmetry in all transmission channels.  }

\begin{figure}\centering
\includegraphics[scale=0.46]{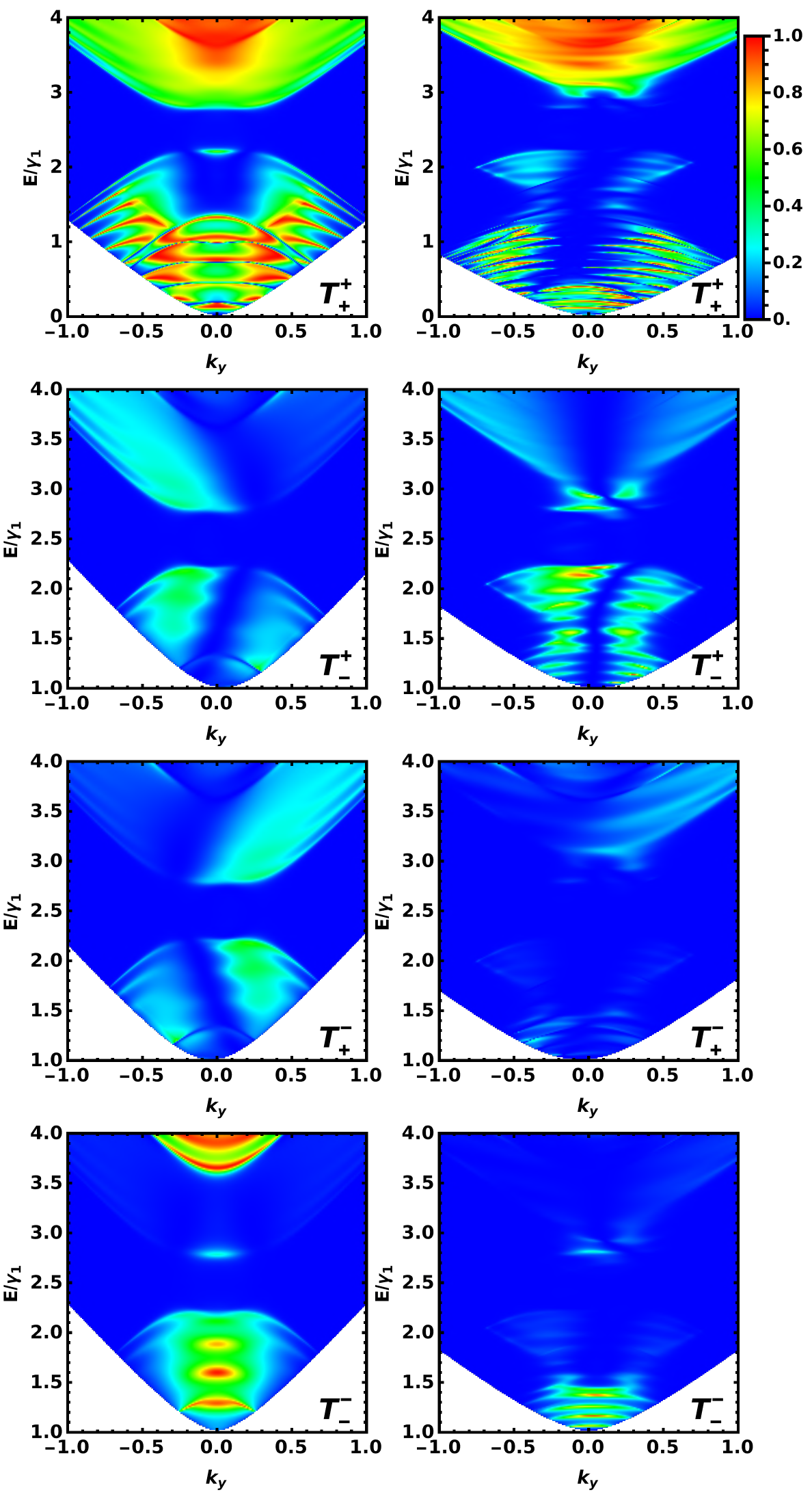}
\caption{(Color online): The same parameters as in Fig. \ref{DensityPlot.pdf}, but now for $\delta_3=0.3\gamma_1$ and $\delta_2=\delta_4=0.$}
\label{DensityDeltas3=0.3.pdf}
\end{figure}
 
Fig. \ref{DensityDeltas3=0.3.pdf} shows the transmission channels for the same parameters as in Fig. \ref{DensityPlot.pdf}, but with $\delta_3=0.3\gamma_1$ and $\delta_2=\delta_4=0$. We observe that the presence of the interlayer bias results in a gap region in all transmission channels, as found in the single barrier case \cite{JELLAL2015149}. In addition, no transmission is found in the gap region, in contrast to \cite{Mouhafid,Lu,Redouani}.
The cloak effect in the $T^+_+$ channel is larger in the triple barrier case than in the single barrier case, as shown in Fig. \ref{DensityPlot.pdf}. However, in the $T^+_-$ channel, it occurs for the same energy region in both the triple barrier and single barrier cases. Furthermore, we observe two thin resonances in the $T^+_-$ channel and triple barrier case that are not present in the single barrier case \cite{JELLAL2015149}. The transmission probability in the $T^-_+$ channel and triple barrier case becomes non-zero in the region of energies greater than $U_4$, in contrast to the result in Fig. \ref{DensityPlot.pdf}, where it is close to zero. However, it still remains lower than in the case of the single barrier. In the $T^-_-$ channel, the transmission in both single barrier and triple barrier cases is similar to the results obtained in Fig. \ref{DensityPlot.pdf}, except that the number of resonances decreases in the single barrier case. 

\subsection{Conductance}

\begin{figure}[H]
	\centering
	\includegraphics[scale=0.7]{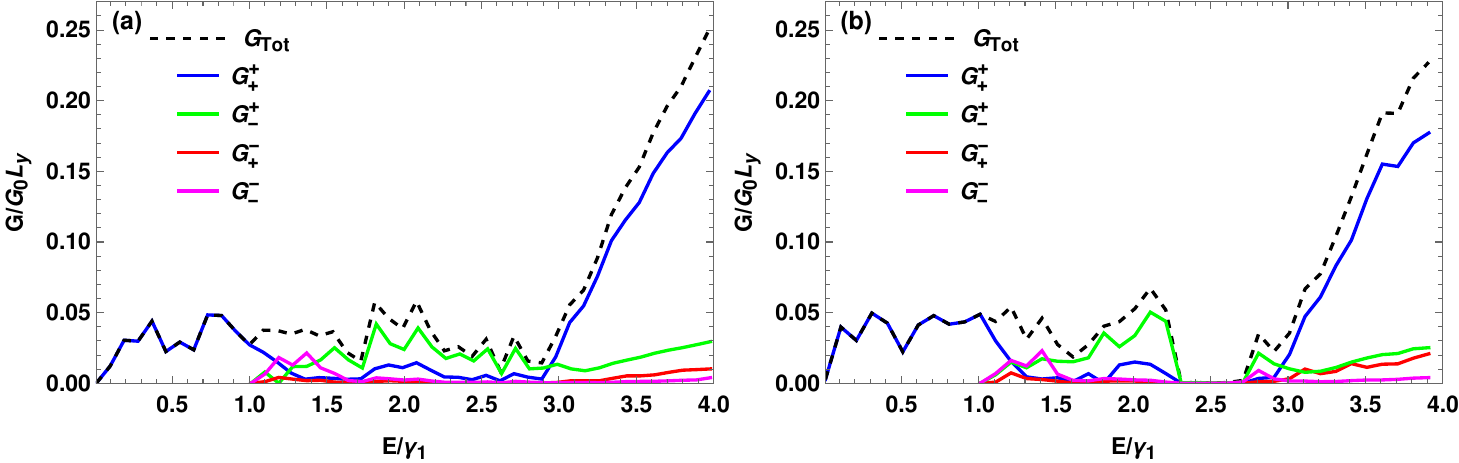}
	\caption{(Color online): Conductance of the triple barrier structure as a function of energy $E$. (a): for $U_2=1.5\gamma_1<U_3=2.5\gamma_1<U_4=3\gamma_1$ and $\delta_2=\delta_3=\delta_4=0$. (b): The same parameters as in \ref{Conductance} (a) but now for $\delta_3=0.3 \gamma_1$ and $\delta_2=\delta_4=0$. The barrier
		widths are set at $b_1= b_2=b_3=15$ nm and $l_B=13.5$ nm.}
	\label{Conductance}
\end{figure}

{In Fig. \ref{Conductance}, we plot the conductance as a function of energy $E$ for the triple barrier system $U_2<U_3<U_4$. Fig. \ref{Conductance} (a) shows the conductance for zero interlayer bias: $\delta_2=\delta_3=\delta_4=0$, which corresponds to the results in Fig. \ref{DensityPlot.pdf} (right). For energy $E<\gamma_1$ , the only contribution to the total conductance comes from $G^+_+$ ($G_{\text{tot}}=G^+_+$), while the other contributions are zero. This is explained by the presence of the $k^+$ propagation mode in the $T^+_+$ channel in this region, whereas it is not present in the other channels, see Fig. \ref{DensityPlot.pdf} (right panel). In the energy region $\gamma_1<E<U_4$, the conductance $G^+_+$ decreases and approaches zero around the points $E=U_2$ and $E=U_3$. This reduction is due to the low transmission and the magnitude of the cloak effect within this region in $T^+_+$ channel. However, $G^+_-$ begins to conduct at $E=\gamma_1$ and provides an important contribution, preventing the total conductance from dropping  to zero in the energy region  $\gamma_1<E<U_4$. This is the result of the large probability of transmission in $T^+_-$ channel, in contrast to the single barrier case \cite{JELLAL2015149}, where the transmission is low in the $T^+_-$ channel, and then it causes the conductance total to drop to zero at $E=U_3$. Furthermore, the contribution from $G^-_-$ in the region $\gamma_1<E<U_2$, where there is a propagation mode $k^-$ in the $T^-_-$ channel (Fig. \ref{DensityPlot.pdf},right panel) and the contribution from $G^+_+$ in the region $U_2<E<U_3$ have enhanced the total conductance $G_{\text{tot}}$. The contribution of $G^-_+$ remains very close to zero until $E=U_4$ due to the large magnitude of the cloak effect in $T^-_+$. Note that for $E>U_4$, $G^+_+$ provides a very strong contribution, which pushes up the total conductance. In Fig. \ref{Conductance} (b), we plot the conductance with the same parameters as in Fig. \ref{Conductance} (a), but now for $\delta_3=0.3\gamma_1$, which corresponds to the transmission channels in Fig. \ref{DensityDeltas3=0.3.pdf} (right panel). The presence of the interlayer bias results in  zero conductance in the gap region for all channels of conductance ($G_{\text{tot}}=G^+_+=G^+_-=G^-_+=G^-_-=0$). For energy $E<\gamma_1$, additional peaks are found in the conductance $G_{\text{tot}}$, which are absent in Fig. \ref{Conductance} (a). In addition, in contrast to Fig. \ref{Conductance}. (a) $G^-_+$ makes a modest contribution in region $\gamma_1<E<U_2$, resulting in a new peak at $E=U_2-\delta_3$ in the total conductance $G_{\text{tot}}$. For $E>U_4$, its contribution increases and approaches $G^+_-$. However, near the gap, we note the suppression of a pick in $G^+_-$ when compared with the result in \ref{Conductance} (a). This suppression is also observed in the total conductance $G_{\text{tot}}$. }

\section{Conclusion}\label{CC}

In this paper, we have investigated the transmission of charge carriers in AB bilayer graphene through a triple potential barrier and in the presence of a perpendicular magnetic field. 
{Thanks to the presence of the magnetic field, we have examined different configurations of the triple barrier in the case of two-band tunnels. It emerged that transmission increases in the case of ascending barriers ($U_2<U_4$) compared to declining barriers ($U_2>U_4$). Additionally, anti-Klein tunneling depends on the value of $U_2$. In the presence of a magnetic field, the application of an interlayer bias in the central region (region 3) is sufficient to open a gap for some configurations of the triple barrier. At fixed energy, the triple barrier enhances transmission when compared to the single barrier case.}

In the four-band tunneling, it was seen that the transmission decreases in the channels $T^+_+$,  $T^-_+$ and  $T^-_-$ when compared to the case of a single barrier. By contrast, in the $T^+_-$ channel, there is an increase in the transmission probability in comparison with the single barrier case. 
{Asymmetry is found in all channels of transmission as a result of the presence of the magnetic field}. Contrary to \cite{Mouhafid,Lu,Redouani}, we observed total suppression of transmission in the gap region in the presence of an interlayer bias, $(\delta_3=0.3\gamma_1)$. {This results in a region of zero conductance (gap) in the total conductance $G_{\text{tot}}$. Additional peaks are found in $G_{\text{tot}}$ in the region before the gap, while near this region a pick is suppressed.} 
To sum up, our findings revealed interesting oscillatory features that provide good regulation of charge carrier transmission in AB bilayer graphene. Our results, on the other hand, demonstrated that in the presence of a magnetic field, the triple barrier system is capable of confining electrons in AB bilayer graphene as well as providing a logic state on/off to graphene-based transistors.

\end{document}